\documentclass[prd,preprint,superscriptaddress,showpacs,byrevtex]{revtex4}
\usepackage{bm}
\def\fsl#1{\setbox0=\hbox{$#1$}           
   \dimen0=\wd0                                 
   \setbox1=\hbox{/} \dimen1=\wd1               
   \ifdim\dimen0>\dimen1                        
      \rlap{\hbox to \dimen0{\hfil/\hfil}}      
      #1                                        
   \else                                        
      \rlap{\hbox to \dimen1{\hfil$#1$\hfil}}   
      /                                         
   \fi}                                         %
\newcommand{\be}{\begin{equation}}
\newcommand{\ee}{\end{equation}}
\newcommand{\bea}{\begin{eqnarray}}
\newcommand{\eea}{\end{eqnarray}}
\newcommand{\beq}{\begin{equation}}
\newcommand{\eeq}{\end{equation}}
\newcommand{\beqs}{\begin{eqnarray}}
\newcommand{\eeqs}{\end{eqnarray}}

\newcommand{\aslash}{A\hspace{-0.067in}\slash}

\begin{document}
\title{ Correct Definition of The QCD Potential From The Wilson Loop}
\author{Gouranga C Nayak }\thanks{E-Mail: nayakg138@gmail.com}
%
%
\date{\today}
\begin{abstract}
The static Coulomb potential energy between the electron and positron at rest separated by a large distance $R$ obtained from the Wilson loop in QED is same as the Coulomb potential energy obtained in the classical Maxwell theory. Since the Yang-Mills theory was discovered by making analogy with the Maxwell theory by extending U(1) group to the SU(3) group one finds by making analogy with the QED that the QCD potential energy between the quark and antiquark at rest separated by a large distance $R$ obtained from the Wilson loop in QCD is the same potential energy obtained in the classical Yang-Mills theory. This implies that the static QCD potential energy $V(R)$ obtained at the large separation distance $R$ in the literature is not consistent with the classical Yang-Mills theory because the potential energy $V(T,R)$ in the classical Yang-Mills theory is time $T$ dependent even if the quark and antiquark are at rest. In this paper we find the correct definition of the QCD potential energy $V(T,R)$ from the Wilson loop in QCD which, at the large separation distance $R$, is consistent with the classical Yang-Mills theory.
\end{abstract}
\maketitle
\pagestyle{plain}

\pagenumbering{arabic}

\section{Introduction}

The Yang-Mills theory \cite{ym9} was discovered in the year 1954 which is an extension of the U(1) gauge theory [Maxwell theory] to the SU(3) gauge theory. The quantization of the classical Yang-Mills theory led to the quantum chromodynamics (QCD). The QCD is a fundamental theory of the nature which describes the interaction between the quarks and gluons. The quarks and gluons are the fundamental particles of the nature which exist inside the hadron (such as inside the proton and  neutron etc.).

The QCD is a renormalizable theory \cite{tv9}. Due to the asymptotic freedom in the renormalized QCD \cite{gw9} the QCD coupling becomes small at the small distance and becomes large at the large distance. Hence the short distance (high momentum transfer) partonic scattering cross section can be calculated by using the perturbative QCD (pQCD). Using the factorization theorem in QCD \cite{fc9,fc91,fc92} the hadronic cross section at the high energy colliders can be calculated from the partonic cross section by using the experimentally extracted parton distribution function (PDF) and fragmentation function (FF).

The hadron formation from the quarks and gluons is a long distance phenomenon in QCD where the coupling becomes large. Hence the pQCD cannot be applied to study the hadron formation from the quarks and gluons. The non-perturbative QCD is necessary to study the hadron formation from quarks and gluons. However, the analytic solution of the non-perturbative QCD is not known yet due to the presence of the cubic and quartic gluon field terms in the QCD lagrangian [see section IV]. Due to this reason the lattice QCD method is used to study the hadron formation from the quarks and gluons.

Note that even if the Yang-Mills theory was discovered in the year 1954 \cite{ym9} the exact form of the Yang-Mills potential $A_\nu^d(x)$ is not known yet where $\nu=0,1,2,3$ is the Lorentz index and $d=1,...,8$ is the color index. This is in contrast to the Maxwell theory where the exact form of the Maxwell potential [the electromagnetic potential] $A_\nu(x)$ is known. Since the exact form of the Coulomb potential [the electric potential] $A_0(x)$ is known it is widely used to study the atomic bound states in the Bohr's atomic model and in the Schrodinger equation. Hence the exact form of the Yang-Mills potential $A_\nu^d(x)$ will be useful to study the bound state hadron formation from the quarks. Because of this reason it is desirable to obtain the exact form of the Yang-Mills potential $A_\nu^d(x)$. Note that the color potential also plays an important role to study the quark-gluon plasma at RHIC and LHC \cite{q9,q91,q92,q93}.

Recently we have found the general form of the color potential [the Yang-Mills potential] $A_\nu^d(x)$ produced by the color charge $q^a(t)$ of the quark \cite{gn9}. We have found that the general form of the color potential [the Yang-Mills potential] $A_\nu^d(x)$ produced by the color charge $q^a(t)$ of the quark at rest is given by \cite{gn9}
\bea
\Phi^d(t,{\vec x})=A_0^d(t,{\vec x})=\frac{q^b(t-r)}{r}\left[\frac{{\rm exp}[g\int dr \frac{Q(t-r)}{r}]-1}{g\int dr \frac{Q(t-r)}{r}}\right]_{db},~~~~Q_{db}(t)=f^{dba}q^a(t),~~~~~{\vec A}^d(t,{\vec x})=0\nonumber \\
\label{pt9}
\eea
where $\int dr$ is an indefinite integration and $r=|{\vec x}-{\vec X}|$ with ${\vec X}$ being the position of the quark at rest. In this paper we use the natural unit.

For constant color charge $q^a$ we find from the above equation $\Phi^a(r)=A_0^a(r)=\frac{q^a}{r}$ which is Coulomb-like potential. However, this Coulomb-like potential $\Phi^a(r)=A_0^a(r)=\frac{q^a}{r}$ is not a potential in the Yang-Mills theory because the constant color charge $q^a$ produces abelian-like color current density $j_\mu^a(x)$ which satisfies the continuity equation $\partial^\nu j_\nu^a(x)=0$ similar to that in abelian-like theory. Since the Yang-Mills color current density $j_\mu^a(x)$ satisfies the equation
\bea
D^\nu[A]j_\nu^b(x)=0,~~~~~~~~~~~~~D_\nu^{bd}[A]=\delta^{bd}\partial_\nu +gf^{bad}A_\nu^a(x)
\label{cu9}
\eea
we find that the color charge $q^a(t)$ of the quark in the Yang-Mills theory is time dependent \cite{gn9,gn91}.

From eq. (\ref{pt9}) we find that the color potential [the Yang-Mills potential] $A_0^a(t,{\vec x})$ produced by the color charge $q^a(t)$ of the quark in the classical Yang-Mills theory is time dependent even if the quark is at rest. This is a consequence of the time dependent color charge $q^a(t)$ of the quark. This implies that the potential energy $V(T,R)$ between the static quark and antiquark separated by a large distance $R$ in the classical Yang-Mills theory is time $T$ dependent even if the quark and antiquark are at rest [see section II for details].

This is in contrast to the present literature where the QCD potential energy $V(R)$ between static quark and antiquark separated by a large distance $R$ obtained by the expectation of the Wilson loop in QCD is defined to be time $T$ independent, {\it i. e.}, it is defined to be $V(R)$ instead of $V(T,R)$.

It should be mentioned here that the Yang-Mills theory was discovered by making analogy with the Maxwell theory by extending U(1) group to SU(3) group \cite{ym9,gn9}. The static Coulomb potential energy between the electron and positron at rest separated by a large distance $R$ obtained from the expectation of the Wilson loop in QED is same as the Coulomb potential energy obtained in the classical Maxwell theory [see section III for details]. Since the Yang-Mills theory was discovered by making analogy with the Maxwell theory by extending U(1) group to the SU(3) group \cite{ym9,gn9} one finds by making analogy with the QED that the QCD potential energy between the quark and antiquark at rest separated by a large distance $R$ obtained from the expectation of the Wilson loop in QCD is the same potential energy obtained in the classical Yang-Mills theory. This implies that the static QCD potential energy $V(R)$ obtained at the large separation distance $R$ in the literature is not consistent with the classical Yang-Mills theory because the potential energy $V(T,R)$ in the classical Yang-Mills theory is time $T$ dependent even if the quark and antiquark are at rest.

In this paper we obtain the correct definition of the QCD potential energy $V(T,R)$ from the expectation of the Wilson loop in QCD which, at the large separation distance $R$, is consistent with the classical Yang-Mills theory. We find that the correct definition of the QCD potential energy $V(T,R)$ at the large separation distance $R$ obtained from the expectation of the Wilson loop in QCD is given by
\bea
V(T,R)=- \frac{d}{dT} {\rm ln}\left[[\frac{<W_C[T'+T,R]>}{<W_C[T',R]>}]_{T'\rightarrow \infty}\right]
\label{ywgl9i}
\eea
where
\bea
W_C[T,R]={\rm Tr}{\cal P}e^{igT^d\oint_C dx^\nu {\hat A}^d_\nu(x)}
\label{ywgh9i}
\eea
is the gauge invariant Wilson loop in QCD \cite{wl9} along the closed path $C$ of spatial extension $R=|{\vec X}_1-{\vec X}_2|$ with ${\vec X}_1$ (${\vec X}_2$) being the position of the quark (antiquark) at rest, $T$ being the temporal extension of the closed path $C$ and ${\hat A}_\nu^d(x)$ is the gluon field. Note that we have used the notation $A_\nu^d(x)$ for the classical Yang-Mills field and the notation ${\hat A}^d_\nu(x)$ for the gluon field (note the hat on the gluon field). In eq. (\ref{ywgl9i}) the expectation of the Wilson loop in QCD is given by
\bea
&& <W_C[T,R]>=\frac{\int [d{\hat A}]   \times W_C[T,R] \times {\rm det}[\frac{\delta \partial^\nu {\hat A}_\nu^b}{\delta \omega^c} ]
\times {\rm exp}[i\int d^4x [-\frac{1}{4} {\hat F}_{\nu \lambda }^d(x) {\hat F}^{\nu \lambda d}(x)-\frac{1}{2\alpha} [\partial^\nu {\hat A}^d_\nu(x)]^2]]}{\int [d{\hat A}] {\rm det}[\frac{\delta \partial^\nu {\hat A}_\nu^b}{\delta \omega^c} ]
\times {\rm exp}[i\int d^4x [-\frac{1}{4} {\hat F}_{\nu \lambda }^d(x) {\hat F}^{\nu \lambda d}(x)-\frac{1}{2\alpha} [\partial^\nu {\hat A}^d_\nu(x)]^2]]}\nonumber \\
\label{ywgm9i}
\eea
where $W_C[T,R]$ is given by eq. (\ref{ywgh9i}), the $\alpha$ is the gauge fixing parameter and
\bea
{\hat F}_{\nu \lambda}^d(x) = \partial_\nu {\hat A}_\lambda^d(x)  - \partial_\lambda {\hat A}_\nu^d(x)  +gf^{dba} {\hat A}_\nu^b(x) {\hat A}_\lambda^a(x)
\label{hfmn9}
\eea
is the non-abelian gluon field tensor. In eq. (\ref{ywgm9i}) we have used the covariant gauge fixing $G_f^a(x)=\partial^\nu {\hat A}^a_\nu(x)$ but it can be done in any arbitrary gauge fixing $G_f^a(x)$.

Hence we find that the time dependent potential energy between static quark and antiquark separated by a large distance is due to the non-zero $f^{abc}$ in the Yang-Mills theory. This can be seen from eqs. (\ref{cu9}) and (\ref{pt9}) as follows. When all the $f^{abc}$ are zero then we find from eq. (\ref{cu9}) that the color charge $q^a$ is constant which gives from eq. (\ref{pt9}) the Coulomb-like potential $\Phi^a(r)=A_0^a(r)=\frac{q^a}{r}$ which is independent of time. Hence the time dependent potential energy between static quark and antiquark separated by a large distance is due to the non-zero $f^{abc}$ in the Yang-Mills theory. This implies that the correct definition of the QCD potential energy at the large separation distance obtained from the Wilson loop in QCD in eq. (\ref{ywgl9i}) must include the $f^{abc}$ term in the non-abelian gluon field tensor ${\hat F}_{\nu \lambda}^d(x) $ in eq. (\ref{hfmn9}) in the path integration in eq. (\ref{ywgm9i}).

In this paper we will provide a derivation of eq. (\ref{ywgl9i}).

The paper is organized as follows. In section II we discuss the time dependent potential energy of the static quark and antiquark separated by a large distance $R$ in the classical Yang-Mills theory. In section III we discuss the QED potential energy from the expectation of the Wilson loop in QED. In section IV we obtain the correct definition of the QCD potential energy $V(T,R)$ from the expectation of the Wilson loop in QCD which, at the large separation distance $R$, is consistent with the classical Yang-Mills theory. Section V contains conclusions.

\section{ Time dependent potential energy of static quark and antiquark separated by a large distance in Yang-Mills theory}

In this section we will discuss the time $T$ dependent potential energy $V(T,R)$ of the static quark and antiquark separated by a large distance $R$ in the classical Yang-Mills theory. Let us first discuss the time independent potential energy $V(R)$ of the static electron and positron separated by a large distance $R$ in the classical Maxwell theory before proceeding to the time dependent potential energy $V(T,R)$ of the static quark and antiquark separated by a large distance $R$ in the classical Yang-Mills theory. Note that although the form of the static potential energy (Coulomb potential energy) in the classical Maxwell theory is well known but we will repeat its derivation here because we will follow the similar procedure for the classical Yang-Mills theory in section IIB.

\subsection{Static potential energy (Coulomb Potential Energy) of electron and positron at rest separated by a large distance in Maxwell theory }

Consider a system in the classical Maxwell theory where the positron at rest is at the position ${\vec X}_1$ having charge $+e$ and the electron at rest is at the position ${\vec X}_2$ having charge $-e$ separated by a large distance $R$ given by
\bea
R=|{\vec X}_1-{\vec X}_2|.
\label{r9}
\eea
The total charge of the system is zero but the charge density is not zero.

The Maxwell equation is given by
\bea
\partial_\mu F^{\mu \lambda}(x)=j^\lambda(x),~~~~~~~~~~~~~~~~~~\partial_\mu F_{\nu \lambda}(x)+\partial_\nu F_{\lambda \mu}(x)+\partial_\lambda F_{\mu \nu}(x)=0
\label{mx9}
\eea
where
\bea
F_{\mu \lambda}(x) =\partial_\mu A_\lambda(x) -\partial_\lambda A_\mu(x)
\label{fm9}
\eea
is the electromagnetic field tensor, $j^\mu(x)$ is the electromagnetic current density and $A^\mu(x)$ is the electromagnetic potential.

Since the electron and positron are at rest in the above system we find
\bea
{\vec j}(x)=0,~~~~~~~~~{\vec A}(x)=0,~~~~~~~~~~~~~~~{\vec B}(x)=0
\label{j9}
\eea
where ${\vec B}(x)$ is the magnetic field. From eqs. (\ref{mx9}), (\ref{fm9}) and (\ref{j9}) we find
\bea
{\vec \nabla}\cdot {\vec E}(x) = \rho(x)=j_0(x),~~~~~~~~~~~~~{\vec \nabla}\times {\vec E}(x)=0,~~~~~~~~~~~~~{\vec E}(x) =-{\vec \nabla}A_0(x)
\label{ne9}
\eea
where $\rho(x)$ is the charge density which is non-zero.

The total interaction energy in this system is given by
\bea
V(t)=\frac{1}{2}\int d^3x \rho(t,{\vec x}) A_0(t,{\vec x}) = \frac{1}{2} \int d^3x {\vec E}(t,{\vec x}) \cdot {\vec E}(t,{\vec x}) + \frac{1}{2}\int d^3x {\vec \nabla} \cdot [{\vec E}(t,{\vec x}) A_0(t,{\vec x})].
\label{ra9}
\eea
The zeroth component of the electromagnetic potential $A_0(t,{\vec x})$ at any position ${\vec x}$ in this system is given by
\bea
A_0(t,{\vec x}) = \frac{e}{|{\vec x}-{\vec X}_1|}-\frac{e}{|{\vec x}-{\vec X}_2|}.
\label{a09}
\eea
From eq. (\ref{ne9}) we find
\bea
\rho(x) =-{\vec \nabla}^2 A_0(x)
\label{rh9}
\eea
which by using eq. (\ref{a09}) gives
\bea
\rho(t,{\vec x}) = e \delta^{(3)}( {\vec x}-{\vec X}_1) -e \delta^{(3)}( {\vec x}-{\vec X}_2).
\label{rh19}
\eea
Since $A_0(t,{\vec x})$ in eq. (\ref{a09}) is time independent and $\rho(t,{\vec x})$ in eq. (\ref{rh19}) is time independent we find from eq. (\ref{ra9}) that the total interaction energy in this system is time independent which is given by
\bea
V=\frac{1}{2}\int d^3x \rho(t,{\vec x}) A_0(t,{\vec x}).
\label{tpe9}
\eea
By using eqs. (\ref{a09}) and (\ref{rh19}) in (\ref{tpe9}) we find [by neglecting the infinite self interaction energies] that
\bea
V(R) =-\frac{e^2}{R},~~~~~~~~~~~~R=|{\vec X}_1-{\vec X}_2|
\label{cpe9}
\eea
which is the static potential energy (Coulomb potential energy) between the electron and positron at rest separated by a large distance $R$ in the classical Maxwell theory.

\subsection{ Time dependent potential energy of static quark and antiquark separated by a large distance in Yang-Mills theory}

Consider a system in the classical Yang-Mills theory where the quark is at rest at the position ${\vec X}_1$ and the antiquark is at rest at the position ${\vec X}_2$ separated by a large distance $R=|{\vec X}_1-{\vec X}_2|$.

The Yang-Mills equation is given by
\bea
D_\nu[A] F^{\nu \lambda b}(x)=j^{\lambda b}(x),~~~~~~~~D_\mu[A] F_{\nu \lambda}^b(x)+D_\nu[A] F_{\lambda \mu}^b(x)+D_\lambda[A] F_{\mu \nu }^b(x)=0
\label{ymx9}
\eea
where
\bea
F_{\mu \lambda}^b(x) =\partial_\mu A_\lambda^b(x) -\partial_\lambda A_\mu^b(x)+gf^{bcd} A_\mu^c(x)A_\lambda^d(x)
\label{yfm9}
\eea
is the Yang-Mills field tensor, $j_\mu^b(x)$ is the color current density, $A_\mu^b(x)$ is the Yang-Mills potential [the color potential] and $D_\nu^{bd}[A]$ is the covariant derivative given by eq. (\ref{cu9}).

Since the quark and antiquark are at rest in the above system we find from eq. (\ref{yfm9}) and (\ref{pt9}) that
\bea
{\vec A}^d(x)=0,~~~~~~~~~~~~~~~{\vec B}^d(x)=0
\label{yj9}
\eea
where ${\vec B}^a(x)$ is the chromo-magnetic field. From eqs. (\ref{ymx9}), (\ref{yfm9}) and (\ref{yj9}) we find
\bea
{\vec \nabla}\cdot {\vec E}^d(x) = \rho^d(x)=j_0^d(x),~~~~~~~~~~~~~{\vec \nabla}\times {\vec E}^d(x)=0,~~~~~~~~~~~~~{\vec E}^d(x) =-{\vec \nabla}A_0^d(x)
\label{yne9}
\eea
where ${\vec E}^a(x)$ is the chromo-electric field and $\rho^a(x)=j_0^a(x)$ is the color charge density.

From eq. (\ref{pt9}) we find for the quark and antiquark at rest by using ${\vec \nabla}\cdot {\vec E}^a(x) = \rho^a(x)=j_0^a(x)$ as given by eq. (\ref{yne9}) that
\bea
&& {\vec E}^d(t,{\vec x})={\hat r}_1 \frac{q^b(t-r_1)}{r^2_1}\left[\frac{{\rm exp}[g\int dr_1 \frac{Q(t-r_1)}{r_1}]-1}{g\int dr_1 \frac{Q(t-r_1)}{r_1}}\right]_{db}
-\frac{\hat r_1}{r_1}\frac{dq^b(t-r_1)}{dr_1}\left[\frac{{\rm exp}[g\int dr_1 \frac{Q(t-r_1)}{r_1}]-1}{g\int dr_1 \frac{Q(t-r_1)}{r_1}}\right]_{db}\nonumber \\
&&-\frac{{\hat r_1}}{r^2_1} q^a(t-r_1)
\left[\left[\frac{{\rm exp}[g\int dr_1 \frac{Q(t-r_1)}{r_1}]-1}{g\int dr_1 \frac{Q(t-r_1)}{r_1}}\right]_{dc}gf^{cap} + \left[\frac{{\rm exp}[g\int dr_1 \frac{Q(t-r_1)}{r_1}]-1}{g\int dr_1 \frac{Q(t-r_1)}{r_1}}\right]_{ca}gf^{cdp} \right]\nonumber \\
&& \left[\frac{1}{g\int dr_1 \frac{Q(t-r_1)}{r_1}}\right]_{pb}q^b(t-r_1)\nonumber \\
&& -{\hat r}_2\frac{q^b(t-r_2)}{r^2_2}\left[\frac{{\rm exp}[g\int dr_2 \frac{Q(t-r_2)}{r_2}]-1}{g\int dr_2 \frac{Q(t-r_2)}{r_2}}\right]_{db}
+\frac{\hat r_2}{r_2}\frac{dq^b(t-r_2)}{dr_2}\left[\frac{{\rm exp}[g\int dr_2 \frac{Q(t-r_2)}{r_2}]-1}{g\int dr_2 \frac{Q(t-r_2)}{r_1}}\right]_{db}\nonumber \\
&&+\frac{{\hat r_2}}{r^2_2} q^a(t-r_2)
\left[\left[\frac{{\rm exp}[g\int dr_2 \frac{Q(t-r_2)}{r_2}]-1}{g\int dr_2 \frac{Q(t-r_2)}{r_2}}\right]_{dc}gf^{cap} + \left[\frac{{\rm exp}[g\int dr_2 \frac{Q(t-r_2)}{r_2}]-1}{g\int dr_2 \frac{Q(t-r_2)}{r_2}}\right]_{ca}gf^{cdp} \right]\nonumber \\
&& \left[\frac{1}{g\int dr_2 \frac{Q(t-r_2)}{r_2}}\right]_{pb}q^b(t-r_2)
\label{cel29}
\eea
where
\bea
r_1=|{\vec x} -{\vec X}_1|,~~~~~~~~~~~r_2=|{\vec x} -{\vec X}_2|.
\label{r1r2}
\eea
Similarly for the quark and antiquark at rest we find from eq. (\ref{cel29}) by using ${\vec E}^a(x) = -{\vec \nabla}A_0^a(x)$ as given by eq. (\ref{yne9}) that
\bea
&& A_0^d(t,{\vec x})=\frac{q^b(t-r_1)}{r_1}\left[\frac{{\rm exp}[g\int dr_1 \frac{Q(t-r_1)}{r_1}]-1}{g\int dr_1 \frac{Q(t-r_1)}{r_1}}\right]_{db} -\frac{q^b(t-r_2)}{r_2}\left[\frac{{\rm exp}[g\int dr_2 \frac{Q(t-r_2)}{r_2}]-1}{g\int dr_2 \frac{Q(t-r_2)}{r_2}}\right]_{db}.\nonumber \\
\label{pt129}
\eea
Note that $\int dr_1$ and $\int dr_2$ integrations are indefinite integrations in eqs. (\ref{cel29}) and (\ref{pt129}).

Similar to eq. (\ref{ra9}) the total interaction energy in this system of quark and antiquark at rest is given by
\bea
V(t)=\frac{1}{2}\int d^3x \rho^d(t,{\vec x}) A_0^d(t,{\vec x}) = \frac{1}{2} \int d^3x {\vec E}^d(t,{\vec x}) \cdot {\vec E}^d(t,{\vec x}) +\frac{1}{2} \int d^3x {\vec \nabla} \cdot [{\vec E}^d(t,{\vec x}) A_0^d(t,{\vec x})].\nonumber \\
\label{yra9}
\eea
Since ${\vec E}^a(t,{\vec x})$ in eq. (\ref{cel29}) is time dependent and $A_0^a(t,{\vec x})$ in eq. (\ref{pt129}) is time dependent we find that $ \int d^3x {\vec E}^a(t,{\vec x}) \cdot {\vec E}^a(t,{\vec x}) + \int d^3x {\vec \nabla} \cdot [{\vec E}^a(t,{\vec x}) A_0^a(t,{\vec x})]$ is time dependent which implies from eq. (\ref{yra9}) that $V(t)$ is time dependent.

Similar to the total interaction energy in eq. (\ref{ra9}) in the Maxwell theory which contains the infinite self interaction energies [see eqs. (\ref{tpe9}) and (\ref{cpe9})], the total interaction energy in eq. (\ref{yra9}) contains the infinite self interaction energies. Since the total interaction energy in eq. (\ref{yra9}) is time dependent we find by neglecting the infinite self energies that the potential energy $V(T,R)$ between the static quark and antiquark separated by a large distance $R$ is time $T$ dependent in the classical Yang-Mills theory.

\subsection{ Conservation Of Energy Is Not Violated Due To Time Dependent Potential Energy of Static Quark and Antiquark in Yang-Mills Theory }

From the Yang-Mills equation we find from eqs. (\ref{cel29}) and (\ref{pt129}) for the quark and antiquark at rest that
\bea
\frac{\partial {\vec E}^d(t,{\vec x})}{\partial t} = gf^{dcb} A_0^c(t,{\vec x}){\vec E}^b(t,{\vec x}) + {\vec j}^d(t,{\vec x}).
\label{encv9}
\eea
From eq. (\ref{encv9}) we find
\bea
{\vec j}^b(t,{\vec x}) \neq 0,~~~~~~~~~{\rm for~the~quark~and~antiquark~at~rest}
\label{vcc9}
\eea
which gives from eq. (\ref{encv9})
\bea
\frac{d[\frac{1}{2} \int d^3x {\vec E}^b(t,{\vec x}) \cdot {\vec E}^b(t,{\vec x})]}{dt} =\int d^3x {\vec j}^b(t,{\vec x}) \cdot {\vec E}^b(t,{\vec x}).
\label{encv19}
\eea
Eq. (\ref{encv19}) is the statement of the conservation of energy in the Yang-Mills theory for the quark and antiquark at rest where $\int d^3x {\vec j}^b(t,{\vec x}) \cdot {\vec E}^b(t,{\vec x})$ is the rate of work done which is non-zero even if the quark and antiquark are rest.

Hence we find that the time dependent potential energy of the static quark and antiquark does not violate the conservation of energy in the Yang-Mills theory.

\section{ QED Potential Energy From The expectation of The Wilson Loop}

In this section we will discuss the QED potential energy from the expectation of the Wilson loop in QED. Although the derivation of the QED potential energy from the expectation of the Wilson loop in QED is well known but we will present its derivation here because we will follow the similar steps in the derivation of the QCD potential energy from the expectation of the Wilson loop in QCD in the next section.

The Wilson loop in QED is given by
\bea
U_C[T,R]=e^{-ie\oint_C dx^\nu {\hat A}_\nu(x)}
\label{wqe9}
\eea
where ${\hat A}_\mu(x)$ is the photon field and $C$ is a closed path of spatial extension $R$ and temporal extension $T$. Note that we have used the notation $A_\nu(x)$ for the classical electromagnetic field and the notation ${\hat A}_\nu(x)$ for the photon field (note the hat on the photon field). The expectation of the Wilson loop is given by
\bea
<U_C[T,R]>=<e^{-ie\oint_C dx^\mu {\hat A}_\mu(x)}> =\frac{\int [d{\hat A}] e^{-ie\oint_C dx^\mu {\hat A}_\mu(x)}\times e^{i\int d^4x [-\frac{1}{4} {\hat F}_{\nu \mu}(x) {\hat F}^{\nu \mu}(x) -\frac{1}{2\alpha} [\partial^\nu {\hat A}_\nu(x)]^2]} }{ \int [d{\hat A}] e^{i\int d^4x [-\frac{1}{4} {\hat F}_{\nu \mu}(x) {\hat F}^{\nu \mu}(x) -\frac{1}{2\alpha} [\partial^\nu {\hat A}_\nu(x)]^2]}}\nonumber \\
\label{vwe9}
\eea
where $\alpha$ is the gauge fixing parameter and
\bea
{\hat F}_{\nu \lambda}(x) = \partial_\nu {\hat A}_\lambda(x)  - \partial_\lambda {\hat A}_\nu(x).
\label{hafmn9}
\eea
In eq. (\ref{vwe9}) we have used the covariant gauge fixing $G_f(x)=\partial^\nu {\hat A}_\nu(x)$ but it can be done in any arbitrary gauge fixing $G_f(x)$.

The eq. (\ref{vwe9}) can be written as
\bea
<U_C[T,R]> =<e^{-ie\oint_C dx^\mu {\hat A}_\mu(x)}> =\frac{\int [d{\hat A}] e^{i\int d^4x [-\frac{1}{4} F_{\nu \mu}(x) F^{\nu \mu}(x) -\frac{1}{2\alpha} [\partial^\nu {\hat A}_\nu(x)]^2 +j^\nu(x) {\hat A}_\nu(x)]} }{ \int [d{\hat A}] e^{i\int d^4x [-\frac{1}{4} F_{\nu \mu}(x) F^{\nu \mu}(x) -\frac{1}{2\alpha} [\partial^\nu {\hat A}_\nu(x)]^2]}}
\label{vwe19}
\eea
where
\bea
j^\nu(x) = e \oint_C dX^\nu \delta^{(4)}(x-X)
\label{vwe29}
\eea
which satisfies the continuity equation
\bea
\partial_\nu j^\nu(x) = 0.
\label{vwe39}
\eea
Note that since the current density in eq. (\ref{vwe29}) satisfies the continuity equation as given by eq. (\ref{vwe39}) one finds that the current density in eq. (\ref{vwe29}) is an admissible current density in Maxwell theory which implies that the interaction term $\int d^4x j^\mu(x) A_\mu(x)$ in eq, (\ref{vwe19}) is the correct interaction action of the photon with the external current density $j^\mu(x)$.

Hence the expectation of the Wilson loop in QED $<e^{-ie\oint_C dx^\mu {\hat A}_\mu(x)}>$ in eq. (\ref{vwe19}) correctly predicts the interaction energy of the photons with the external current density $j^\mu(x)$ in eq. (\ref{vwe29}).

By performing the path integration in eq. (\ref{vwe19}) we find in the Euclidean time
\bea
{\rm ln}<U_C[T,R]> = -\frac{1}{2}\int d^4x j^\nu(x)\frac{1}{\partial^2}[g_{\nu \mu}+(\alpha-1)\frac{1}{\partial^2}\partial_\nu \partial_\mu]j^\mu(x)
\label{vwe49}
\eea
which by using the continuity equation from eq. (\ref{vwe39}) gives
\bea
{\rm ln}<U_C[T,R]> =-\frac{1}{2} \int d^4x j^\nu(x)\frac{1}{\partial^2}j_\nu(x).
\label{vwe59}
\eea
From eq. (\ref{vwe29}) we find
\bea
j^\nu(x) = e \oint_C dT \frac{dX^\nu}{dT} \delta^{(4)}(x-X)=\left[\int_{C_1} dT - \int_{C_2} dT \right] ~e~\frac{dX^\nu}{dT} ~ \delta^{(4)}(x-X)
\label{vwe69}
\eea
where $C_1$ and $C_2$ are two paths which enclose the closed path $C$.

For $\frac{d{\vec X}}{dT}=0$ we find from eq. (\ref{vwe69}) that
\bea
j^\nu(x)=\delta_{\nu 0} e \delta^{(3)}( {\vec x}-{\vec X}_1) - \delta_{\nu 0}e \delta^{(3)}( {\vec x}-{\vec X}_2)
\label{vwe99}
\eea
which is same as eq. (\ref{rh19}). By using eq. (\ref{vwe99}) in (\ref{vwe59}) we find [by neglecting the infinite self interacting energies] that
\bea
V(R)=-\frac{1}{T}{\rm ln}<U_C[T,R]> =-\frac{e^2}{R},~~~~~~~~~~~~R=|{\vec X}_1-{\vec X}_2|
\label{vwe91}
\eea
which is the Coulomb potential energy between the electron and position separated by a large distance $R$ which agrees with the corresponding result in the classical Maxwell theory where $<U_C[T,R]>$ is given by eq. (\ref{vwe9}).

\subsection{QED Potential Energy From Gauge Invariant Green's Function in QED }

Consider the gauge invariant operator in QED given by
\bea
{\cal O}(X_1,X_2)={\bar \psi}(X_2) U_{C_1}[X_2,X_1] \psi(X_1)
\label{gi9}
\eea
where $\psi(x)$ is the Dirac field of the fermion in QED and $U_{C_1}[X_1,X_2]$ is the Wilson line given by
\bea
U_{C_1}[X_2,X_1]=e^{-ie\int_{X_1}^{X_2} dx^\nu {\hat A}_\nu(x)}
\label{wll9}
\eea
where $C_1$ is the path joining the points $X_1$ and $X_2$.

Let us evaluate the vacuum expectation of the gauge invariant correlation function of the type $<0|{\cal O}(X_1,X_2){\cal O}(X_3,X_4)|0>$ in QED where the gauge invariant ${\cal O}(X_1,X_2)$ is given by eq. (\ref{gi9}) and $|0>$ is the vacuum state of the full QED. We find
\bea
&& <0|{\cal O}(X_1,X_2){\cal O}(X_3,X_4)|0>=<0|{\bar \psi}(X_2) U_{C_1}[X_2,X_1] \psi(X_1){\bar \psi}(X_4) U_{C_2}[X_4,X_3] \psi(X_3)|0>\nonumber \\
&&=\frac{1}{Z[0]}\int [d{\hat A}][d{\bar \psi}] [d\psi] {\bar \psi}(X_2) U_{C_1}[X_2,X_1] \psi(X_1){\bar \psi}(X_4) U_{C_2}[X_4,X_3] \psi(X_3) {\rm exp}[i\int d^4x [-\frac{1}{4} {\hat F}_{\nu \mu}(x) {\hat F}^{\nu \mu}(x) \nonumber \\
&&-\frac{1}{2\alpha} [\partial^\nu {\hat A}_\nu(x)]^2+{\bar \psi}(x)[i{\not \partial}-m -e{\hat \aslash}(x)]\psi(x)]]
\label{wla9}
\eea
which can be written as
\bea
&& <0|{\bar \psi}(X_2) U_{C_1}[X_2,X_1] \psi(X_1){\bar \psi}(X_4) U_{C_2}[X_4,X_3] \psi(X_3)|0>\nonumber \\
&&=\frac{1}{Z[0]}\int [d{\hat A}]\frac{\delta}{\delta \eta(X_2)}  U_{C_1}[X_2,X_1] \frac{\delta}{\delta {\bar \eta}(X_1)} \frac{\delta}{\delta \eta(X_4)}  U_{C_2}[X_4,X_3] \frac{\delta}{\delta {\bar \eta}(X_3)}{\rm exp}[i\int d^4x [-\frac{1}{4} {\hat F}_{\nu \mu}(x) {\hat F}^{\nu \mu}(x) \nonumber \\
&&-\frac{1}{2\alpha} [\partial^\nu {\hat A}_\nu(x)]^2]] \times \int [d{\bar \psi}] [d\psi]  {\rm exp}[i\int d^4x [
{\bar \psi}(x)[i{\not \partial}-m -e{\hat \aslash}(x)]\psi(x)+ {\bar \eta}(x) \cdot \psi(x)\nonumber \\
&&+ {\bar \psi}(x) \cdot \eta(x)]]|_{\eta={\bar \eta}=0}
\label{wlb9}
\eea
where $Z[0]$ is the generating functional in QED in the absence of any external sources.

By change of variables we find
\bea
&& <0|{\bar \psi}(X_2) U_{C_1}[X_2,X_1] \psi(X_1){\bar \psi}(X_4) U_{C_1}[X_4,X_3] \psi(X_3)|0>\nonumber \\
&&=\frac{1}{Z[0]}\int [d{\hat A}]\frac{\delta}{\delta \eta(X_2)}  U_{C_1}[X_2,X_1] \frac{\delta}{\delta {\bar \eta}(X_1)} \frac{\delta}{\delta \eta(X_4)}  U_{C_2}[X_4,X_3] \frac{\delta}{\delta {\bar \eta}(X_3)}\nonumber \\
&&\times {\rm exp} [i\int d^4x \int d^4y {\bar \eta}(x)S(x,y,{\hat A})\eta(y)] \times {\rm exp}[i\int d^4x [-\frac{1}{4} {\hat F}_{\nu \mu}(x) {\hat F}^{\nu \mu}(x)-\frac{1}{2\alpha} [\partial^\nu {\hat A}_\nu(x)]^2]] \nonumber \\
&&\times \int [d{\bar \psi}] [d\psi]  {\rm exp}[i\int d^4x [
{\bar \psi}(x)[i{\not \partial}-m -e{\hat \aslash}(x)]\psi(x)]]|_{\eta={\bar \eta}=0}
\label{wlc9}
\eea
where the $S(x,y,{\hat A})$ is given by
\bea
[i{\not \partial}-m -e{\hat \aslash}(x)]S(x,y,{\hat A})=\delta^{(4)}(x-y).
\label{wld9}
\eea
By performing the gaussian integration of the fermion fields in eq. (\ref{wlc9}) we find
\bea
&& <0|{\bar \psi}(X_2) U_{C_1}[X_2,X_1] \psi(X_1){\bar \psi}(X_4) U_{C_2}[X_4,X_3] \psi(X_3)|0>=\frac{1}{Z[0]}\int [d{\hat A}]{\rm Tr}[S(X_3,X_2,{\hat A})  U_{C_1}[X_2,X_1]\nonumber \\ &&S(X_1,X_4,{\hat A})  U_{C_2}[X_4,X_3]] \times {\rm exp}[i\int d^4x [-\frac{1}{4} {\hat F}_{\nu \mu}(x) {\hat F}^{\nu \mu}(x)-\frac{1}{2\alpha} [\partial^\nu {\hat A}_\nu(x)]^2]]\times {\rm det}[i{\not \partial}-m -e{\hat \aslash}(x)].\nonumber \\
\label{wle9}
\eea
Since we will be considering the potential energy at the large separation distance the fermion loop contributions are small at the large distance. Hence we will put ${\rm det}[i{\not \partial}-m -e{\hat \aslash}(x)]=1$ in eq. (\ref{wle9}) which gives
\bea
&& <0|{\bar \psi}(X_2) U_{C_1}[X_2,X_1] \psi(X_1){\bar \psi}(X_4) U_{C_2}[X_4,X_3] \psi(X_3)|0>=\frac{1}{Z'[0]}\int [d{\hat A}]{\rm Tr}[S(X_3,X_2,{\hat A})  U_{C_1}[X_2,X_1] \nonumber \\
&& S(X_1,X_4,{\hat A})  U_{C_2}[X_4,X_3] ] \times {\rm exp}[i\int d^4x [-\frac{1}{4} {\hat F}_{\nu \mu}(x) {\hat F}^{\nu \mu}(x)-\frac{1}{2\alpha} [\partial^\nu {\hat A}_\nu(x)]^2]]
\label{wlf9}
\eea
where $Z'[0]$ is obtained from $Z[0]$ by putting ${\rm det}[i{\not \partial}-m -e{\hat \aslash}(x)]=1$.

From eq. (\ref{wld9}) we find
\bea
i\partial_0 S(x,y,{\hat A})+i{\vec \alpha}\cdot {\vec \nabla}S(x,y,{\hat A})=\gamma^0\delta^{(4)}(x-y)+[e{\hat A}_0(x)-e{\vec \alpha} \cdot {\hat {\vec A}}(x)+\gamma^0m]S(x,y,{\hat A})\nonumber \\
\label{sla9}
\eea
which can be solved as follows. The solution of the partial differential equation
\bea
A \partial_0 f+B\partial_1 f + C\partial_2 f + D \partial_3 f = H
\label{dif19}
\eea
is given by
\bea
\frac{dt}{A} = \frac{dx_1}{B}=\frac{dx_2}{C}=\frac{dx_3}{D}=\frac{df}{H}=K
\label{dif29}
\eea
which can be seen by comparing with the equation
\bea
dt \partial_0 f+dx_1\partial_1 f + dx_2\partial_2 f + dx_3 \partial_3 f= df.
\label{dif39}
\eea
Hence we find from eq. (\ref{sla9}) that
\bea
i\frac{dS(x,y,{\hat A})}{dt} = \gamma^0\delta^{(4)}(x-y)+[e\frac{dx^\mu}{dt} {\hat A}_\mu (x)+\gamma^0m]S(x,y,{\hat A}).
\label{eq3}
\eea
For the electron and positron at rest separated by a large distance the constant mass $m$ term contributes to a constant part of the energy so that it does not contribute to the potential energy. This implies that we can drop the mass $m$ term in eq. (\ref{eq3}) to obtain the potential energy between the electron and positron at rest separated by a large distance. Hence by dropping the mass $m$ term in eq. (\ref{eq3}) we find
\bea
i\frac{dS(x,y,{\hat A})}{dt} -e\frac{dx^\mu}{dt} {\hat A}_\mu (x)S(x,y,{\hat A})= \gamma^0\delta^{(4)}(x-y).
\label{eq4}
\eea
Since eq. (\ref{eq4}) is an ordinary differential equation we find the solution
\bea
i S(X,Y,{\hat A})=U_{C_1}[{\vec X},T;{\vec Y},T]~\gamma^0\delta^{(3)}({\vec X}-{\vec Y})
\label{eq8}
\eea
where $U_{C_1}[{\vec X}_1,T;{\vec X}_2,T]$ is the Wilson line in QED given by eq. (\ref{wll9}). Using eq. (\ref{eq8}) in (\ref{wlf9}) we find
\bea
&& <0| {\cal O}^\dagger(T,R) {\cal O}(0,R)|0>=\frac{N}{Z'[0]}\int [d{\hat A}]   U_C[T,R]
\times {\rm exp}[i\int d^4x [-\frac{1}{4} {\hat F}_{\nu \mu}(x) {\hat F}^{\nu \mu}(x)-\frac{1}{2\alpha} [\partial^\nu {\hat A}_\nu(x)]^2]]\nonumber \\
\label{wgg9}
\eea
where
\bea
U_C[T,R]=e^{-ie\oint_C dx^\mu {\hat A}_\mu(x)}
\label{wgh9}
\eea
is the Wilson loop in QED with $C$ being the closed path of spatial (temporal) extension $R$ ($T$) and $N$ is a factor which is not important as it will cancel in the ratio in eq. (\ref{wgl9}).

Inserting a complete set of states we find in the Euclidean time
\bea
&& <0| {\cal O}^\dagger(T,R) {\cal O}(0,R)|0>= \sum_n |<0|{\cal O}|H_n>|^2 e^{-TE_n(R)}
\label{wgi9}
\eea
which for large time $T\rightarrow \infty$ gives
\bea
&& [<0| {\cal O}^\dagger(T,R) {\cal O}(0,R)|0>]_{T\rightarrow \infty}= |<0|{\cal O}|H>|^2 e^{-TE(R)}.
\label{wgj9}
\eea
For electron and positron at rest we find
\bea
&& [<0| {\cal O}^\dagger(T,R) {\cal O}(0,R)|0>]_{T\rightarrow \infty}= |<0|{\cal O}|H>|^2 e^{-TV(R)}.
\label{wgk9}
\eea
Hence from eqs. (\ref{wgk9}) and (\ref{wgg9}) we find
\bea
V(R)=- \frac{d}{dT} {\rm ln}\left[[\frac{<U_C[T'+T,R]>}{<U_C[T',R]>}]_{T'\rightarrow \infty}\right]
\label{wgl9}
\eea
where
\bea
&& <U_C[T,R]>=\frac{\int [d{\hat A}] U_C[T,R]
\times {\rm exp}[i\int d^4x [-\frac{1}{4} {\hat F}_{\nu \mu}(x) {\hat F}^{\nu \mu}(x)-\frac{1}{2\alpha} [\partial^\nu {\hat A}_\nu(x)]^2]]}{\int [d{\hat A}]
\times {\rm exp}[i\int d^4x [-\frac{1}{4} {\hat F}_{\nu \mu}(x) {\hat F}^{\nu \mu}(x)-\frac{1}{2\alpha} [\partial^\nu {\hat A}_\nu(x)]^2]]}.
\label{wgm9}
\eea
Since the QED potential energy $V(R)$ of static electron and positron separated by a large distance $R$ is independent of time $T$ we find from eq. (\ref{wgl9}) that
\bea
V(R)=- \frac{1}{T} {\rm ln}<U_C[T,R]>|_{T\rightarrow \infty}
\label{wgl9ff}
\eea
which reproduces eq. (\ref{vwe91}).

\section{ Correct Definition of The QCD Potential From The Wilson Loop}

The Wilson loop in QCD is given by
\bea
W_C[T,R]={\rm Tr}{\cal P}e^{igT^d\oint_C dx^\nu {\hat A}^d_\nu(x)}
\label{ywqe9}
\eea
where ${\hat A}^a_\mu(x)$ is the gluon field and $C$ is the closed path of spatial extension $R$ and temporal extension $T$. Similar to the photon case in eq. (\ref{vwe9}) the expectation of the Wilson loop for the gluon case is given by
\bea
&& <W_C[T,R]>=<{\rm Tr}{\cal P}e^{igT^d\oint_C dx^\nu {\hat A}^d_\nu(x)}> \nonumber \\
&&=\frac{\int [d{\hat A}][{\rm Tr}{\cal P}e^{igT^a\oint_C dx^\mu {\hat A}^a_\mu(x)}] \times {\rm det}[\frac{\delta \partial^\nu {\hat A}_\nu^b}{\delta \omega^c} ] \times {\rm exp}[i\int d^4x [-\frac{1}{4} {\hat F}_{\nu \mu }^a(x) {\hat F}^{\nu \mu a}(x) -\frac{1}{2\alpha} [\partial^\nu {\hat A}^a_\nu(x)]^2]]}{\int [d{\hat A}] {\rm det}[\frac{\delta \partial^\nu {\hat A}_\nu^b}{\delta \omega^c} \times {\rm exp}[i\int d^4x [-\frac{1}{4} {\hat F}_{\nu \mu }^a(x) {\hat F}^{\nu \mu a}(x) -\frac{1}{2\alpha} [\partial^\nu {\hat A}^a_\nu(x)]^2]]}\nonumber \\
\label{yvwe9}
\eea
where $\alpha$ is the gauge fixing parameter and ${\hat F}_{\nu \mu }^a(x)$ is the non-abelian gluon field tensor given by eq. (\ref{hfmn9}).

From eq. (\ref{vwe9}) one finds that the current density $j_\mu(x)$ in eq. (\ref{vwe29}) satisfies the continuity equation (\ref{vwe39}) which means the current density in eq. (\ref{vwe29}) is an admissible current density in the Maxwell theory. Similarly, from eq. (\ref{yvwe9}), if one defines the color current density of the form
\bea
j_\mu^a(x) = gT^a \oint_C dy_\mu \delta^{(4)}(x-y)
\label{yvwe29}
\eea
then this color current density is not an admissible color current density in the Yang-Mills theory because the $T^a$ is a matrix but the color current density $j_\mu^a(x)$ in the Yang-Mills theory should not be a matrix. Note that the trace is taken in eqs. (\ref{ywqe9}) and (\ref{yvwe9}) but the trace of $j_\mu^a(x)$ is zero in eq. (\ref{yvwe29}) which means it is not straightforward to find an interaction action of the form $\int d^4x j_\mu^a(x) A^{\mu a}(x)$ from eq. (\ref{yvwe9}) for the gluon case although it was easy for the photon case in eq. (\ref{vwe19}) in the Maxwell theory.

Hence it is not straightforward to prove that the expectation $<{\rm Tr}{\cal P}e^{igT^a\oint_C dx^\mu {\hat A}^a_\mu(x)}>$ of the Wilson loop in QCD in eq. (\ref{yvwe9}) correctly predicts the interaction energy in QCD.

In this section we will obtain the correct definition of the QCD potential energy between quark and antiquark separated by a large distance from the expectation $<{\rm Tr}{\cal P}e^{igT^a\oint_C dx^\mu {\hat A}^a_\mu(x)}>$ of the Wilson loop in QCD.

\subsection{ Correct Definition of The QCD Potential Energy From The Wilson Loop }

Consider the gauge invariant operator in QCD
\bea
{\cal O}(X_1,X_2)={\bar \psi}(X_2) W_{C_1}[X_2,X_1] \psi(X_1)
\label{ygi9}
\eea
where $\psi(x)$ is the Dirac field of the quark and $W_{C_1}[X_1,X_2]$ is the Wilson line in QCD given by
\bea
W_{C_1}[X_2,X_1]={\cal P}e^{igT^d\int_{X_1}^{X_2} dx^\nu {\hat A}^d_\nu(x)}
\label{ywll9}
\eea
where $C_1$ is the path joining the points $X_1$ and $X_2$.

Let us evaluate the vacuum expectation of the gauge invariant correlation function of the type $<0|{\cal O}(X_1,X_2){\cal O}(X_3,X_4)|0>$ in QCD where the gauge invariant operator ${\cal O}(X_1,X_2)$ is given by eq. (\ref{ygi9}) and $|0>$ is the vacuum state of the full QCD. The suppression of color indices are understood.

We find
\bea
&& <0|{\cal O}(X_1,X_2){\cal O}(X_3,X_4)|0>=<0|{\bar \psi}(X_2) W_{C_1}[X_2,X_1] \psi(X_1){\bar \psi}(X_4) W_{C_2}[X_4,X_3] \psi(X_3)|0>\nonumber \\
&&=\frac{1}{Z[0]}\int [d{\hat A}][d{\bar \psi}] [d\psi] {\bar \psi}(X_2) W_{C_1}[X_2,X_1] \psi(X_1){\bar \psi}(X_4) W_{C_2}[X_4,X_3] \psi(X_3) \times {\rm det}[\frac{\delta \partial^\nu {\hat A}_\nu^b}{\delta \omega^c} ] \nonumber \\
&& \times {\rm exp}[i\int d^4x [-\frac{1}{4} {\hat F}_{\nu \mu}^a(x) {\hat F}^{\nu \mu a}(x) -\frac{1}{2\alpha} [\partial^\nu {\hat A}^a_\nu(x)]^2+{\bar \psi}(x)[i{\not \partial}-m+ gT^a{\hat \aslash}^a(x)]\psi(x)]]
\label{ywla9}
\eea
which gives
\bea
&& <0|{\bar \psi}(X_2) W_{C_1}[X_2,X_1] \psi(X_1){\bar \psi}(X_4) W_{C_2}[X_4,X_3] \psi(X_3)|0>\nonumber \\
&&=\frac{1}{Z[0]}\int [d{\hat A}]\frac{\delta}{\delta \eta(X_2)}  W_{C_1}[X_2,X_1] \frac{\delta}{\delta {\bar \eta}(X_1)} \frac{\delta}{\delta \eta(X_4)}  W_{C_2}[X_4,X_3] \frac{\delta}{\delta {\bar \eta}(X_3)}\nonumber \\
&&\times  {\rm det}[\frac{\delta \partial^\nu {\hat A}_\nu^b}{\delta \omega^c} ]\times {\rm exp}[i\int d^4x [-\frac{1}{4} {\hat F}_{\nu \mu}^a(x) {\hat F}^{\nu \mu a}(x) -\frac{1}{2\alpha} [\partial^\nu {\hat A}^a_\nu(x)]^2]] \times \int [d{\bar \psi}] [d\psi] \nonumber \\
&& {\rm exp}[i\int d^4x [
{\bar \psi}(x)[i{\not \partial}-m+ gT^a{\hat \aslash}^a(x)]\psi(x)+ {\bar \eta}(x) \cdot \psi(x)
+ {\bar \psi}(x) \cdot \eta(x)]]|_{\eta={\bar \eta}=0}
\label{ywlb9}
\eea
where $Z[0]$ is the generating functional in QCD in the absence of any external sources.

By change of variables we find
\bea
&& <0|{\bar \psi}(X_2) W_{C_1}[X_2,X_1] \psi(X_1){\bar \psi}(X_4) W_{C_2}[X_4,X_3] \psi(X_3)|0>\nonumber \\
&&=\frac{1}{Z[0]}\int [d{\hat A}]\frac{\delta}{\delta \eta(X_2)}  W_{C_1}[X_2,X_1] \frac{\delta}{\delta {\bar \eta}(X_1)} \frac{\delta}{\delta \eta(X_4)}  W_{C_2}[X_4,X_3] \frac{\delta}{\delta {\bar \eta}(X_3)}\times {\rm det}[\frac{\delta \partial^\nu {\hat A}_\nu^b}{\delta \omega^c} ]\nonumber \\
&&\times  {\rm exp} [i\int d^4x \int d^4y {\bar \eta}(x)S(x,y,{\hat A})\eta(y)]  \times {\rm exp}[i\int d^4x [-\frac{1}{4} {\hat F}_{\nu \mu}^a(x) {\hat F}^{\nu \mu a}(x)-\frac{1}{2\alpha} [\partial^\nu {\hat A}^a_\nu(x)]^2]] \nonumber \\
&&\times \int [d{\bar \psi}] [d\psi]  {\rm exp}[i\int d^4x [
{\bar \psi}(x)[i{\not \partial}-m+ gT^a{\hat \aslash}^a(x)]\psi(x)]]|_{\eta={\bar \eta}=0}
\label{ywlc9}
\eea
where the $S(x,y,{\hat A})$ is given by
\bea
[i{\not \partial}-m+ gT^a{\hat \aslash}^a(x)]S(x,y,{\hat A})=\delta^{(4)}(x-y).
\label{ywld9}
\eea
By performing the gaussian integration of the fermion fields in eq. (\ref{ywlc9}) we find
\bea
&& <0|{\bar \psi}(X_2) W_{C_1}[X_2,X_1] \psi(X_1){\bar \psi}(X_4) W_{C_2}[X_4,X_3] \psi(X_3)|0>\nonumber \\
&&=\frac{1}{Z[0]}\int [d{\hat A}]{\rm Tr}[S(X_3,X_2,{\hat A})  W_{C_1}[X_2,X_1] S(X_1,X_4,{\hat A})  W_{C_2}[X_4,X_3]] \times {\rm det}[\frac{\delta \partial^\nu {\hat A}_\nu^b}{\delta \omega^c} ]\nonumber \\
&& \times {\rm exp}[i\int d^4x [-\frac{1}{4} {\hat F}_{\nu \mu}^a(x) {\hat F}^{\nu \mu a}(x)-\frac{1}{2\alpha} [\partial^\nu {\hat A}^a_\nu(x)]^2]]\times {\rm det}[i{\not \partial}-m+ gT^a{\hat \aslash}^a(x)].
\label{ywle9}
\eea
Since we will be considering the potential energy at the large separation distance the fermion loop contributions are small at the large distance. Hence we will put ${\rm det}[i{\not \partial}-m+ gT^a{\hat \aslash}^a(x)]=1$ in eq. (\ref{ywle9}) which gives
\bea
&& <0|{\bar \psi}(X_2) W_{C_1}[X_2,X_1] \psi(X_1){\bar \psi}(X_4) W_{C_2}[X_4,X_3] \psi(X_3)|0>\nonumber \\
&&=\frac{1}{Z'[0]}\int [d{\hat A}]{\rm Tr}[S(X_3,X_2,{\hat A})  W_{C_1}[X_2,X_1] S(X_1,X_4,{\hat A})  W_{C_2}[X_4,X_3] ] \nonumber \\
&&\times {\rm det}[\frac{\delta \partial^\nu {\hat A}_\nu^b}{\delta \omega^c} ]\times {\rm exp}[i\int d^4x [-\frac{1}{4} {\hat F}_{\nu \mu}^a(x) {\hat F}^{\nu \mu a}(x)-\frac{1}{2\alpha} [\partial^\nu {\hat A}^a_\nu(x)]^2]]
\label{ywlf9}
\eea
where $Z'[0]$ is obtained from $Z[0]$ by putting ${\rm det}[i{\not \partial}-m+ gT^a{\hat \aslash}^a(x)]=1$.

From eq. (\ref{ywld9}) we find
\bea
i\partial_0 S(x,y,{\hat A})+i{\vec \alpha}\cdot {\vec \nabla}S(x,y,{\hat A})=\gamma^0\delta^{(4)}(x-y)-[gT^a{\hat A}^a_0(x)-gT^a{\vec \alpha} \cdot {\hat {\vec A}}^a(x)+\gamma^0m]S(x,y,{\hat A}).\nonumber \\
\label{ysla9}
\eea
For the quark and antiquark at rest separated by a large distance the constant mass $m$ term contributes to a constant part of the energy so that it does not contribute to the potential energy. This implies that we can drop the mass $m$ term in eq. (\ref{ysla9}) to obtain the potential energy between the quark and antiquark at rest separated by a large distance. Hence by dropping the mass $m$ term in eq. (\ref{ysla9}) we find
\bea
i\partial_0 S(x,y,{\hat A})+i{\vec \alpha}\cdot {\vec \nabla}S(x,y,{\hat A})=\gamma^0\delta^{(4)}(x-y)-[gT^a{\hat A}^a_0(x)-gT^a{\vec \alpha} \cdot {\hat {\vec A}}^a(x)]S(x,y,{\hat A}).\nonumber \\
\label{ysla9a}
\eea
which can be solved as follows.

The solution of the partial differential equation in eq. (\ref{dif19}) can be obtained from eq. (\ref{dif29}) [see eq. (\ref{dif39})] which means we find from eq. (\ref{ysla9a})
\bea
i\frac{dS(x,y,{\hat A})}{dt} +gT^a\frac{dx^\mu}{dt} {\hat A}^a_\mu (x)S(x,y,{\hat A})= \gamma^0\delta^{(4)}(x-y).
\label{yeq4}
\eea
Since eq. (\ref{yeq4}) is an ordinary differential equation [similar to the eq. (\ref{eq4}) in QED] we find from eq. (\ref{yeq4}) the solution
\bea
iS(X,Y,{\hat A})=W_{C_1}[{\vec X},T;{\vec Y},T]~\gamma^0\delta^{(3)}({\vec X}-{\vec Y})
\label{yeq8}
\eea
where $W_{C_1}[{\vec X},T;{\vec Y},T]$ is the Wilson line in QCD given by eq. (\ref{ywll9}). Using eq. (\ref{yeq8}) in (\ref{ywlf9}) we find
\bea
&& <0| {\cal O}^\dagger(T,R) {\cal O}(0,R)|0>=\frac{N'}{Z'[0]}\int [d{\hat A}]   W_C[T,R]\times {\rm det}[\frac{\delta \partial^\nu {\hat A}_\nu^b}{\delta \omega^c} ] \nonumber \\
&& \times {\rm exp}[i\int d^4x [-\frac{1}{4} {\hat F}_{\nu \mu }^d(x) {\hat F}^{\nu \mu d}(x)-\frac{1}{2\alpha} [\partial^\nu {\hat A}_\nu(x)]^2]]
\label{yywgg9}
\eea
where
\bea
W_C[T,R]={\rm Tr}{\cal P}e^{igT^d\oint_C dx^\nu {\hat A}^d_\nu(x)}
\label{ywgh9}
\eea
is the Wilson loop in QCD with $C$ being the closed path of spatial (temporal) extension $R$ ($T$) and $N'$ is a factor which is not important as it will cancel in the ratio in eq. (\ref{ywgl9}).

Inserting a complete set of states we find in the Euclidean time
\bea
&& <0| {\cal O}^\dagger(T,R) {\cal O}(0,R)|0>= \sum_n |<0|{\cal O}|H_n>|^2 e^{-\int dTE_n(T,R)}
\label{ywgi9}
\eea
which for the large time $T\rightarrow \infty$ gives
\bea
&& [<0| {\cal O}^\dagger(T,R) {\cal O}(0,R)|0>]_{T\rightarrow \infty}= |<0|{\cal O}|H>|^2 e^{-\int dTE(T,R)}
\label{ywgj9}
\eea
where $\int dT$ is an indefinite integration. For quark and antiquark at rest we find
\bea
&& [<0| {\cal O}^\dagger(T,R) {\cal O}(0,R)|0>]_{T\rightarrow \infty}= |<0|{\cal O}|H>|^2 e^{-\int dTV(T,R)}.
\label{ywgk9}
\eea
Hence from eqs. (\ref{ywgk9}) and (\ref{yywgg9}) we find
\bea
V(T,R)=- \frac{d}{dT} {\rm ln}\left[[\frac{<W_C[T'+T,R]>}{<W_C[T',R]>}]_{T'\rightarrow \infty}\right]
\label{ywgl9}
\eea
which is the correct definition of the QCD potential energy $V(T,R)$ between static quark and antiquark separated by a large distance $R$ obtained from the Wilson loop in QCD where $<W_C[T,R]>$ is given by eq. (\ref{ywgm9i}). The eq. (\ref{ywgl9}) reproduces eq. (\ref{ywgl9i}).

\section{Conclusions}
The static Coulomb potential energy between the electron and positron at rest separated by a large distance $R$ obtained from the Wilson loop in QED is same as the Coulomb potential energy obtained in the classical Maxwell theory. Since the Yang-Mills theory was discovered by making analogy with the Maxwell theory by extending U(1) group to the SU(3) group one finds by making analogy with the QED that the QCD potential energy between the quark and antiquark at rest separated by a large distance $R$ obtained from the Wilson loop in QCD is the same potential energy obtained in the classical Yang-Mills theory. This implies that the static QCD potential energy $V(R)$ obtained at the large separation distance $R$ in the literature is not consistent with the classical Yang-Mills theory because the potential energy $V(T,R)$ in the classical Yang-Mills theory is time $T$ dependent even if the quark and antiquark are at rest. In this paper we have found the correct definition of the QCD potential energy $V(T,R)$ from the Wilson loop in QCD which, at the large separation distance $R$, is consistent with the classical Yang-Mills theory.


\begin{thebibliography}{99}

\bibitem{ym9} C. N. Yang and R. Mills, Phys. Rev. 96 (1954) 191.

\bibitem{tv9} G. 't Hooft and M.J.G. Veltman, Nucl.Phys. B44 (1972) 189.

\bibitem{gw9} D. J. Gross and F. Wilczek, Phys. Rev. Lett. 30 (1973) 1343; D. Politzer, Phys. Rev. Lett. 30 (1973) 1346.

\bibitem{fc9} J. C. Collins, D. E. Soper and G. Sterman, Nucl. Phys. B261 (1985) 104; G. C. Nayak, J. Qiu and G. Sterman, Phys. Lett. B613 (2005) 45; Phys. Rev. D72 (2005) 114012; Phys. Rev. D74 (2006) 074007; Phys. Rev. D77 (2008) 034022.

\bibitem{fc91} G. T. Bodwin {\it et al.}, arXiv:1910.05497 [hep-ph].

\bibitem{fc92} G. C. Nayak, JHEP 1709 (2017) 090; Eur. Phys. J. C76 (2016) 448; Eur. Phys. J. Plus 133 (2018) 52; Phys. Part. Nucl. Lett. 13 (2016) 417; arXiv:1506.02651 [hep-ph]; Phys. Part. Nucl. Lett. 14 (2017) 18; J. Theor. Appl. Phys. 11 (2017) 275.

\bibitem{q9} G. C. Nayak and P. van Nieuwenhuizen, Phys. Rev. D 71 (2005) 125001; G. C. Nayak {\it et al.}, Nucl. Phys. A687 (2001) 457; F. Cooper, C-W. Kao and G. C. Nayak, Phys. Rev. D66 (2002) 114016; G. C. Nayak, Phys. Lett. B442 (1998) 427; JHEP 9802 (1998) 005.

\bibitem{q91} M. C. Birse, C-W. Kao and G. C. Nayak, Phys. Lett. B570 (2003) 171; C-W. Kao, G. C. Nayak and W. Greiner, Phys. Rev. D66 (2002) 034017; A. Chamblin, F. Cooper and G. C. Nayak, Phys. Rev. D69 (2004) 065010; Phys. Lett. B672 (2009) 147; Phys. Rev. D70 (2004) 075018.

\bibitem{q92} F. Cooper, E. Mottola and G. C. Nayak, Phys. Lett. B555 (2003) 181; D. Dietrich, G. C. Nayak and W. Greiner, Phys. Rev. D64 (2001) 074006; G. C. Nayak and R. S. Bhalerao, Phys. Rev. C 61 (2000) 054907; G. C. Nayak and V. Ravishankar, Phys. Rev. C 58 (1998) 356; Phys. Rev. D 55 (1997) 6877.

\bibitem{q93} G. C. Nayak, Phys. Rev. D 72 (2005) 125010; Eur. Phys. J. C59 (2009) 715; Annals Phys. 325 (2010) 682; Eur. Phys. J.C59 (2009) 891; Eur. Phys. J. C64 (2009) 73; JHEP 0906 (2009) 071; Annals Phys. 324 (2009) 2579; Annals Phys. 325 (2010) 514.

\bibitem{gn9} G. C. Nayak, JHEP 1303 (2013) 001.

\bibitem{gn91} G. C. Nayak, Eur. Phys. J. C73 (2013) 2442.

\bibitem{wl9} K. G. Wilson, Phys. Rev. D10 (1974) 2445.

\end{thebibliography}
\end{document}